\documentclass[article,twocolumn,showpacs,aps,prb,floatfix,superscriptaddress,longbibliography]{revtex4-1}
\usepackage{graphicx,subfigure,epsfig,verbatim,psfrag,amsmath,amssymb,color}
\usepackage{indentfirst}
\usepackage{textcomp}
\usepackage{latexsym}
\usepackage{amssymb}
\usepackage{amsmath}
\usepackage{lipsum}
\usepackage{soul}
\usepackage{xcolor}
\usepackage{color}
\usepackage{epstopdf}
\usepackage{placeins}
\usepackage{float}

\usepackage{xcolor}
\usepackage[urlcolor=blue]{hyperref}      
\hypersetup{
    colorlinks = true,                    
    citecolor = {blue},
    linkcolor = {purple},
           }
\newcommand{\be}{\begin{equation}}
\newcommand{\ee}{\end{equation}}
\newcommand{\bea}{\begin{eqnarray}}
\newcommand{\eea}{\end{eqnarray}}

\begin{document}		
\title{Visualizing Quantum Phases And Identifying Quantum Phase Transitions By Nonlinear Dimensionality Reduction}
\author{Yuan Yang}\thanks{These two authors have contributed equally to this work}
\affiliation{School of Physical Sciences, University of Chinese Academy of Sciences, P. O. Box 4588, Beijing 100049, China}
\author{Zheng-Zhi Sun} \thanks{These two authors have contributed equally to this work}
\affiliation{School of Physical Sciences, University of Chinese Academy of Sciences, P. O. Box 4588, Beijing 100049, China}

\author{Shi-Ju Ran}\email[Corresponding author. ] {Email: sjran@cnu.edu.cn}
\affiliation{Department of Physics, Capital Normal University, Beijing 100048, China}
\author{Gang Su}
\email[Corresponding author. ] {Email: gsu@ucas.ac.cn}
\affiliation{Kavli Institute for Theoretical Sciences, and CAS Center for Excellence in Topological Quantum Computation, University of Chinese Academy of Sciences, Beijing 100190, China}
\affiliation{School of Physical Sciences, University of Chinese Academy of Sciences, P. O. Box 4588, Beijing 100049, China}
\date{\today}

\begin{abstract}
    Identifying quantum phases and phase transitions is key to understand complex phenomena in statistical physics. In this work, we propose an unconventional strategy to access quantum phases and phase transitions by visualization based on the distribution of ground states in Hilbert space. By mapping the quantum states in Hilbert space onto a two-dimensional feature space using an unsupervised machine learning method, distinct phases can be directly specified and quantum phase transitions can be well identified. Our proposal is benchmarked on  gapped, critical, and topological phases in several strongly correlated spin systems. As this proposal directly learns quantum phases and phase transitions from the distributions of the quantum states, it does not require priori knowledge of order parameters of physical systems, which thus indicates a perceptual route to identify quantum phases and phase transitions particularly in complex systems by visualization through learning.
    \\

\end{abstract}
\maketitle

\section{Introduction}
Studying quantum phases and phase transitions in many-body systems belongs to the most challenging topics in contemporary physics. The characterization of quantum phases within Landau paradigm~\cite{landau1937theory, ginzburg2009theory} often requires certain prior knowledge of order parameters. For the phases beyond the Landau paradigm~\cite{wen1989vacuum}, one of the main challenges is how to find proper ``order parameters'', which might be nonlocal or could not be represented by any observables, to characterize the quantum phases~\cite{chen2010local}. While conventional approaches usually rely on the priori knowledge and human wisdom, machine learning (ML) provides an alternative way by training an ML model based on certain given relevant data.
One popular strategy is to utilize a machine learning model (such as neural network or Boltzmann machine) as the classifier to identify the phases of many-body systems \cite{carleo2017solving, carrasquilla2017machine, broecker2017machine, ch2017machine}. In these cases, a supervised learning process is usually involved.

Another promising direction is to incorporate with the unsupervised learning schemes to reveal the phases. For instance, Wang \emph{et al} uses a linear dimensionality reduction algorithm known as principle component analyses (PCA) \cite{pearson1901liii} to identify the thermodynamic phase transitions of classical Ising models \cite{wang2016discovering}. One advantage of the unsupervised learning schemes is that less prior knowledge is required, such as the knowledge on the number of phases and the data for training the model \cite{wang2016discovering,wetzel2017unsupervised,hu2017discovering,wang2017machine,ch2018unsupervised,khatami2020visualizing,van2017learning,rodriguez-nieva2019identifying,zhang2019machine,scheurer2020unsupervised,long2020unsupervised,che2020topological}. However, the unsupervised learning of quantum phases are particularly challenging, mainly due to the exponentially large Hilbert space. In the previous works, one usually implements Monte Carlo samplings to solve this issue \cite{wang2016discovering,wetzel2017unsupervised,hu2017discovering,wang2017machine,ch2018unsupervised,khatami2020visualizing}. It is strongly desired to develop novel, efficient, and simple schemes to learn quantum phases via an unsupervised process.

\begin{figure}[tbp]
	\includegraphics[width=0.95\linewidth]{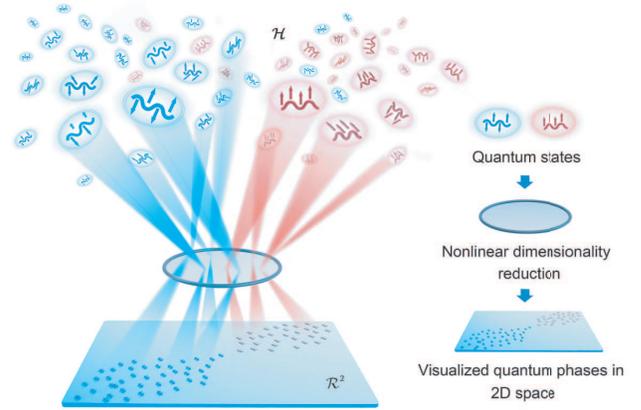}
	\caption{Visualizing quantum states via an unsupervised nonlinear dimensionality reduction: an illustration. The ground states in the Hilbert space $\mathcal{H}$ of exponentially large dimensionality can be mapped onto a two-dimensional feature space $\mathcal{R}^2$ via $t$-SNE, taking negative logarithmic fidelity to measure the distance between two quantum states.
Different points in the feature space represent the ground states at different physical parameters in the Hamiltonian.
The quantum phases and phase transitions can be clearly specified through the visualized distribution of the states in $\mathcal{R}^2$.}
	\label{htsne}
\end{figure}

In this work, we propose an unconventional approach to access the physical information of quantum phases and phase transitions of many-body systems. Our proposal is to probe the ground states (GS) distribution in Hilbert space (denoted as $\mathcal{H}$) by the unsupervised nonlinear dimensionality reduction (DR) scheme \cite{lee2007nonlinear, mokbel2013visualizing,tenenbaum2000global,che2020topological,scheurer2020unsupervised,long2020unsupervised} known as $t$-distributed stochastic neighbor embedding ($t$-SNE) \cite{hinton2003stochastic,maaten2008visualizing,ch2018unsupervised,huembeli2018identifying}.
Such a DR algorithm maps the quantum states from $\mathcal{H}$ to a two-dimensional (2D) feature space (denoted as $\mathcal{R}^2$) by stochastically maximizing the similarity between the GS distribution in $\mathcal{H}$ and that in $\mathcal{R}^{2}$. By simply viewing the distribution in $\mathcal{R}^2$ using naked eyes or employing classical algorithms such as $k$-means \cite{kriegel2011density},  we show that the ground states can be readily classified into correct phases and the critical points of quantum phase transitions can be reliably determined. Our proposal is benchmarked on one-dimensional (1D) quantum lattice models, where we visualize that the quantum states in various phases (including gapped, critical, and topological phases) cluster into different patterns in $\mathcal{R}^{2}$, and the phase transitions can be directly specified. Different from the conventional approaches in many-body physics where one usually focuses on order parameters, entanglements and so on, our work poses a new way to view quantum phases from mutual distances between them. In addition, the present proposal works well not only for quantum data like quantum states but also for classical data like image classification. \\

\begin{figure*}[tbp]
	\includegraphics[width=0.95\linewidth]{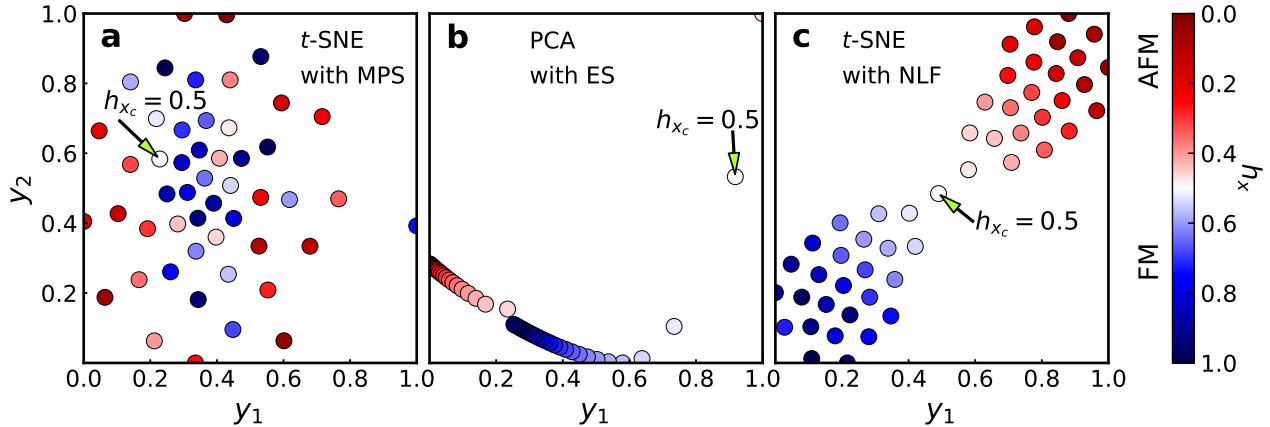}
	\caption{(Color online) The visualizations of distinct quantum phases of the 1D transverse field Ising model in  two-dimensional feature space $\mathcal{R}^2$ for three different measuring distances between quantum states. In {\bf{a}}, the map is implemented by $t$-SNE where the distances of states are measured by the Euclidean distances $D^{\mathcal{M}}_{\alpha, \alpha'} = \| \mathbf{v}^{\alpha} - \mathbf{v}^{\alpha'}  \|$ of the MPS parameters. In {\bf{b}}, we use PCA for mapping and the entanglement spectra \cite{van2017learning} for measuring the distances of states. {\bf{c}} shows the results of our proposal, where the map is implemented by $t$-SNE with negative logarithmic fidelity (NLF) $D^{\mathcal{H}}_{\alpha, \alpha'} = -\log(|\langle\psi^{\alpha} |\psi^{\alpha'}\rangle|)$ as the distance between the ground state. The antiferromagnetic (AFM) and polarized ferromagnetic (FM) phases are clearly specified. Each state is represented by a point on a plane of two components $(y_1, y_2)$ in space $\mathcal{R}^2$, and the color bar indicates the magnitudes of $h_x$.  We calculate 50 GS's at different $h_x$ by DMRG, where the values of $h_x$ are taken uniformly from 0 to 1. For DMRG, we take the size of the system $L=80$ and the dimension cut-off $\chi=30$. For $t$-SNE, we take the number of iteration steps $n_{iter}=5000$ and the perplexity $\mathcal{P}=24$.}
	\label{fig_Ising}
\end{figure*}

\section{Quantum phase visualization}

The central idea of our scheme is to visualize the quantum states by reducing the dimensionality of the exponentially large Hilbert space spanned by the quantum states to two dimensionality using $t$-SNE. The $t$-SNE is a nonlinear DR method that has been widely used in machine learning to visualize high-dimensional data~\cite{maaten2008visualizing}.
Consider a quantum Hamiltonian $\hat{H}(\alpha)$ with $\alpha$ a physical parameter (e.g., a coupling constant or magnetic field), where we suppose a phase transition occurs at $\alpha = \alpha_c$. When $\alpha$ changes continuously, the GS's (denoted as $\{|\psi^{\alpha} \rangle\}$) form a manifold in the Hilbert space $\mathcal{H}$. To proceed, we sample $N$ states by taking $N$ different values of $\alpha$. These quantum states are distributed within the manifold.

Given the data in a high-dimensional space (e.g., the GS's $\{|\psi^{\alpha} \rangle \}$ in the Hilbert space $\mathcal{H}$), one can define the joint probability for each pair of the data ($|\psi^{\alpha} \rangle$ and $|\psi^{\alpha'} \rangle$) as
\begin{equation}
P(D_{\alpha, \alpha'}^{\mathcal{H}}) = \frac{P(\alpha|\alpha') + P(\alpha'|\alpha)}{2N},
\label{eq::Ppair}
\end{equation}
where $N$ is the number of states and the conditional probability is defined by the distances as
\begin{equation}
P(\alpha|\alpha') = \frac{\exp[-(D^{\mathcal{H}}_{\alpha \alpha'})^2/2\sigma_{\alpha}^2]}{\sum_{\beta\neq \alpha} \exp[-(D^{\mathcal{H}}_{\alpha\beta})^2/2\sigma_{\alpha}^2]},
\label{eq::Pij}
\end{equation}
with $\{\sigma_{\alpha}\}$ the hyper-parameters in $t$-SNE, and $D_{\alpha, \alpha'}^{\mathcal{H}}$ is the measure of the distances in $\mathcal{H}$, which can be chosen as different quantities (see below).

One usually does not directly control $\{\sigma_{\alpha}\}$ but define a quantity named as \textit{perplexity} $\mathcal{P}$. Given $\mathcal{P}$, one can perform the binary search to determine $\{\sigma_{\alpha}\}$ that satisfy
\begin{equation}
\log_2 \mathcal{P} = -\sum_{\alpha'}P(\alpha'|\alpha) \log_2 P(\alpha|\alpha'),
\label{eq::perplexity}
\end{equation}
The perplexity controls how non-locally that one state is related to others in the joint probability distributions. More results are provided in the Sec.~\ref{sec::robust} to show the robustness of phase visualization with different perplexities.

To map $\{|\psi^{\alpha} \rangle\}$ onto $\{\mathbf{y}^{\alpha}\}$ in $\mathcal{R}^2$, one can randomly initialize $\{\mathbf{y}^{\alpha}\}$ and define the joint probabilities $\{P(D_{\alpha, \alpha'}^{\mathcal{R}})\}$ as the Student $t$-distribution~\cite{maaten2008visualizing}
\begin{equation}
P(D_{\alpha, \alpha'}^{\mathcal{R}}) = \frac{[1+(D_{\alpha, \alpha'}^{\mathcal{R}})^2]^{-1}}{\sum_{\beta\neq \alpha}[1 + (D_{\alpha, \beta}^{\mathcal{R}})^2]^{-1}},
\label{eq:Qij}
\end{equation}
where the measure of the distances in $\mathcal{R}^2$ is chosen to be the Euclidean distances $D_{\alpha, \alpha'}^{\mathcal{R}} = \|\mathbf{y}^{\alpha} - \mathbf{y}^{\alpha'}\|$.

To capture $\{|\psi^{\alpha} \rangle \}$ by $\{\mathbf{y}^{\alpha}\}$, the strategy of $t$-SNE is to optimize $\{\mathbf{y}^{\alpha}\}$ by minimizing the Kullback-Leibler~(KL) divergence~\cite{kullback1951information} between $\{P(D_{\alpha, \alpha'}^{\mathcal{H}})\}$ and $\{P(D_{\alpha, \alpha'}^{\mathcal{R}})\}$. The KL divergence is defined as
\begin{equation}
\text{KL}(\mathcal{H}, \mathcal{R}^2) = \sum_{\alpha \alpha'} P(D_{\alpha, \alpha'}^{\mathcal{H}}) \log {\frac{P(D_{\alpha, \alpha'}^{\mathcal{H}})} {P(D_{\alpha, \alpha'}^{\mathcal{R}})}}.
\label{eq::KL}
\end{equation}
The gradients by varying $\mathbf{y}^{\alpha}$ are given by
\begin{equation}
\frac{\delta \text{KL}(\mathcal{H}, \mathcal{R}^2)}{\delta \mathbf{y}^{\alpha}} = 4\sum_{\alpha'}  \frac{[P(D_{\alpha, \alpha'}^{\mathcal{H}}) - P(D_{\alpha, \alpha'}^{\mathcal{R}})] (\mathbf{y}^\alpha-\mathbf{y}^{\alpha'})}{1+(D_{\alpha, \alpha'}^{\mathcal{R}})^2}.
\label{eq:partial_C}
\end{equation}
One may use a gradient-descent approach to minimize KL$(\mathcal{H}, \mathcal{R}^2)$. The converged ${\mathbf{y}^\alpha}$ are considered as the embedding of ${\psi^\alpha}$ in $\mathcal{R}^2$ where the mutual distances among $\{|\psi^{\alpha} \rangle \}$ in $\mathcal{H}$ are optimally retained by ${\mathbf{y}^\alpha}$.

To visualize the distribution of certain given GS's $\{|\psi^{\alpha} \rangle \}$, we invoke the recipe of $t$-SNE and map the states onto the vectors $\{\mathbf{y}^{\alpha}\}$ living in a 2D feature space $\mathcal{R}^2$, i.e., $|\psi^{\alpha} \rangle \overset{\text{f}}{\to} \mathbf{y}^{\alpha}$ with $\mathbf{y}^{\alpha} = [y_1^{\alpha}, y_2^{\alpha}]$ a two-component vector and $f$ a nonlinear map from $\mathcal{H}$ to $\mathcal{R}^2$ (Fig.~\ref{htsne}). To be specific, we start from $N$ given states $\{|\psi^{\alpha} \rangle \}$ and define the joint probability distributions $\{P(D_{\alpha, \alpha'}^{\mathcal{H}})\}$ based on the distances $D_{\alpha, \alpha'}^{\mathcal{H}}$ between any two of states $|\psi^{\alpha} \rangle$ and $|\psi^{\alpha'} \rangle$. Then, we randomly initialize $N$ vectors $\{\mathbf{y}^{\alpha}\}$ in $\mathcal{R}^2$ and define joint probability distributions $\{P(D_{\alpha, \alpha'}^{\mathcal{R}})\}$ based on the distances $D_{\alpha, \alpha'}^{\mathcal{R}}$ between any two vectors in $\{\mathbf{y}^{\alpha}\}$. Note that the measure of distance in each space can be chosen flexibly. We choose the Euclidean distance as $D_{\alpha, \alpha'}^{\mathcal{R}}$, and the negative logarithmic fidelity (NLF) \cite{ZOV08fidTN, PhysRevLett.106.055701}
\begin{equation}
D^{\mathcal{H}}_{\alpha, \alpha'} = -\log(|\langle\psi^{\alpha} |\psi^{\alpha'}\rangle|)
\label{eq_NLF}
\end{equation}
to measure the distance between two GS's in $\mathcal{H}$.

To capture the distribution of $\{|\psi^{\alpha} \rangle \}$ by that of $\{\mathbf{y}^{\alpha}\}$, we directly optimize $\{\mathbf{y}^{\alpha}\}$ so that the difference between two probability distributions $\{P(D_{\alpha, \alpha'}^{\mathcal{H}})\}$ and $\{P(D_{\alpha, \alpha'}^{\mathcal{R}})\}$ (averaging over all possible pairs) is minimized. The DR map is left implicit. Consequently, the converged vectors $\{\mathbf{y}^{\alpha}\}$ represent the quantum states $\{|\psi^{\alpha} \rangle \}$
in the 2D feature space of reduced dimensionality.
\\

\section{Identifying Quantum Phases in Spin Models by Visualization}
\subsection{Landau-type phase transition in 1D transverse field Ising spin chain}
We firstly examine our proposal on the 1D transverse field Ising model (TFIM) \cite{LSM61exact}, where the Hamiltonian reads $\hat{H}(h_x) =\sum_{i}\hat{S}_i^z \hat{S}_{i+1}^z - h_x\sum_{i}\hat{S}_i^x$, where $\hat{S}_i^z$ and $\hat{S}_i^x$ stand for the z- and x-component spin operators, respectively, and $h_x$ is the transverse field. It has been rigorously shown that a Landau-type quantum phase transition occurs at the critical field $h_{x}^c=0.5$, which separates the antiferromagnetic (AFM) from polarized ferromagnetic (FM) phases. We employ the density matrix renormalization group (DMRG) \cite{white1992density} to calculate the GS's for different transverse fields in the form of matrix product states (MPS) \cite{fannes1989exact, fannes1992m, perez2006matrix, verstraete2008matrix,ran2020tensor}. The visualizations of quantum phases of the 1D TFIM using three distinct schemes are presented in Fig.~\ref{fig_Ising}. 

In Fig. \ref{fig_Ising} {\bf{a}}, we choose the distance between two GS's as the Euclidean distance $D^{\mathcal{M}}_{\alpha, \alpha'} = \| \mathbf{v}^{\alpha} - \mathbf{v}^{\alpha'}  \|$ for comparison, where the vector $\mathbf{v}^{\alpha}$ is simply formed by all variational parameters in corresponding MPS (i.e. all tensor elements). The $t$-SNE is used to reduce the dimensionality from $\tilde{N}$ to $2$ with $\tilde{N}$ the total number of tensor elements in the MPS. We adopt the canonical form~\cite{OV08canonical} to fix the gauge degrees of freedom of the MPS. It is known that MPS can give an efficient parametrization of the exponentially-large number of GS's of the Hamiltonian under study, where $\tilde{N}$ scales only linearly with the system size~\cite{VC06MPSFaithfully}. However, our results show that the states after DR are mixed up in $\mathcal{R}^2$. It suggests that such a parametrization may not reflect well the quantum state distribution in $\mathcal{H}$. 

In Fig. \ref{fig_Ising} {\bf{b}}, we pick the bipartite entanglement spectra (ES) $\mathbf{s}^{\alpha}$ of the GS's as the input data, which is $\chi$-dimensional with $\chi$ the dimension cut-off in DMRG. Then these ES are mapped onto $\mathcal{R}^2$ by means of PCA \cite{pearson1901liii,wang2016discovering,costa2017principal,wetzel2017unsupervised,hu2017discovering,wang2017machine}.
Different from the $t$-SNE, the PCA uses a linear transformation for DR and obtains the two components in $\mathcal{R}^2$ that optimally retain the covariances of the data in original space.
{PCA succeeds in identifying the phase transitions of the classical spin models based on the sampled spin configurations by
Monte Carlo methods \cite{wang2016discovering,wetzel2017unsupervised,hu2017discovering,wang2017machine}.}
{However, for quantum many-body system, it is hardly to directly input the quantum states to PCA due to its exponentially large Hilbert space.}
{One way for classifying the quantum states by PCA is to input the ES of the quantum states \cite{van2017learning}, where ES is viewed as the effective feature of quantum states.}
The states from the two phases form a 1D stream in $\mathcal{R}^2$ with a break corresponding to the region near the transition point. Our results by PCA are in accordance with those on the Kitaev chain \cite{van2017learning}. As indicated in Fig. \ref{fig_Ising} {\bf{b}}, it is not easy to identify the critical point from the distribution of ES with reduced dimensionality by PCA, {possibly due to the absence of the nonlinearity in the map between $\mathcal{H}$ and $\mathcal{R}^2$.}

Fig. \ref{fig_Ising} {\bf{c}} demonstrates the results using our proposal, in which the $t$-SNE is applied to reduce nonlinearly the dimensionality based on the NLF's [Eq. (\ref{eq_NLF})]. It is obvious that the states inside the AFM and FM phases cluster, and the distribution in $\mathcal{R}^2$ exhibits an ``hourglass'' pattern formed by two oval regions. The critical point between the AFM and FM phases can be easily identified by naked eyes (or by unsupervised learning methods where the two ovals touch each other.
{In previous works, $t$-SNE has been used to classify phases of both the classical spin models \cite{wetzel2017unsupervised} and quantum many-body models \cite{ch2018unsupervised}. Monte Carlo samplings in a given basis are required to obtain the data for implementing DR.
The distances among the quantum states are estimated by Euclidean distances among the sampled (classical) configurations, instead of the states themselves.
In this work, we choose NLF to measure the distances. NLF can be efficiently calculated using TN representation, where Monte Carlo samplings are not required. The relevant stochastic errors in the sampling processes can therefore be avoided.
The result in Fig. \ref{fig_Ising} {\bf{c}} indicates that the NLF is a more proper choice for measuring the distance between two GS's in reducing the dimensionality. The convergence and robustness against small noises of the $t$-SNE with NLF for the visualization of quantum states and phase transitions as well as for the classical data are presented in Sec.~\ref{sec::robust}.

\subsection{Topological-to-magnetic phase transitions in spin-1 chains}

To further demonstrate the ``hourglass'' pattern and the identification of phase transitions by our proposal, we turn to the spin-1 antiferromagnetic Heisenberg uniform chain in a magnetic field ($h_z$), where the Hamiltonian reads $\hat{H}(h_z)=\sum_{i} \sum_{\kappa=x,y,z}  \hat{S}_{i}^{\kappa} \hat{S}_{i+1}^{\kappa} - h_z\sum_{i} \hat{S}_i^z$. For $h_z<h_c$ with the transition point $h_c \simeq 0.414$, the system is in a topological phase known as Haldane phase \cite{haldane1983continuum, haldane1983nonlinear, white1993numerical} with non-trivial boundary excitations and long-rang string orders \cite{nijs1989preroughening,anfuso2007string,anfuso2007fragility}. For $h_z>h_c$, the spin gap is closed by the magnetic filed, and the system enters a topologically trivial magnetic (TTM) phase. As shown in Fig. \ref{fig_HaZz} {\bf{a}}, an ``hourglass''-like distribution emerges, where the Haldane phase and TTM phase are obviously separated. The touching point of the two oval clusters appears at $h_z=0.42$ (note $h_z$ is discretized with the interval $\delta h=0.01$), indicating the critical magnetic field. The estimated critical field by the touching point is slightly higher than expected, possibly due to the finite-size effects that tend to increase the gap (here we take the system size $L=128$ in DMRG).

Fig. \ref{fig_HaZz} {\bf{b}} shows the partten formed by the GS's of the spin-1 Heisenberg AFM model on zigzag chain with nearest neighboring (NN) and next-nearest neighboring (NNN) couplings $\hat{H}(J_1, J_2) = \sum_{i} \sum_{\kappa=x,y,z} (J_1\hat{S}_{i}^{\kappa}\hat{S}_{i+1}^{\kappa} + J_2\hat{S}_{i}^{\kappa} \hat{S}_{i+2}^{\kappa})$, where $J_1$ and $J_2$ denotes the strength of the NN and NNN couplings, respectively. Such a system is frustrated \cite{B10QSLRev} as there is a competition between two kinds of resonating valence bond configurations, of which both possess non-trivial topological properties. A quantum phase transition occurs at $(J_2/ J_1)_c\simeq 0.744$ \cite{kolezhuk1996first}, where the system is in the Haldane phase and the NNN Haldane phase on two sides of the critical point. Again, an ``hourglass'' pattern emerges, where the two phases cluster in  two oval areas. The touching point with $J_2/J_1=0.745$ (the interval of the discretization step $\delta{(J_2/J_1)}=0.005$) accurately identifies the transition point.

\begin{figure}[tbp]
	\includegraphics[width=0.9\linewidth]{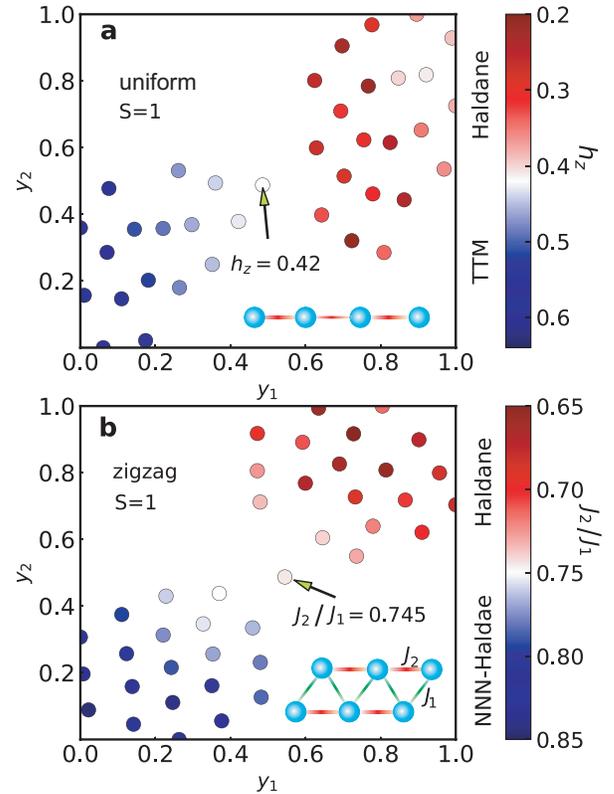}
	\caption{(Color online) The visualization of quantum phases in the spin-1 Heisenberg antiferromagnetic uniform chain {\bf{a}} in different magnetic field $h_z$, and {\bf{b}} the visualization of those of the spin-1 Heisenberg antiferromagnetic zigzag chain with different strength of the next-nearest neighboring (NNN) couplings $J_2/J_1$ (the nearest neighboring coupling is fixed at 1).
In {\bf{a}}, the topologically trivial magnetic (TTM) phase and Haldane phase are clearly specified, where it gives the critical field $h_{z_c} = 0.42$. In {\bf{b}}, the two distinct Haldane phases are obviously separated, which shows the critical NNN coupling $(J_2/J_1)_c = 0.745$.
The distance of quantum states is measured by $D^{\mathcal{H}}_{\alpha, \alpha'} = -\log(|\langle\psi^{\alpha} |\psi^{\alpha'}\rangle|)$.
We take 40 values of $h_z$ and $J_2/J_1$ with the interval $\delta h=0.01$ and $\delta {J_2/J_1}=0.005$ for the two models, respectively. We take the system size $L=128$ and $100$, the dimension cut-off of DMRG $\chi=128$ and $60$ for {\bf{a}} and {\bf{b}}, respectively. Iterative steps in $t$-SNE $n_{iter}=5000$ with the perplexity $\mathcal{P}=20$ for both.}
	\label{fig_HaZz}
\end{figure}

\begin{figure}[tbp]
	\includegraphics[width=0.9\linewidth]{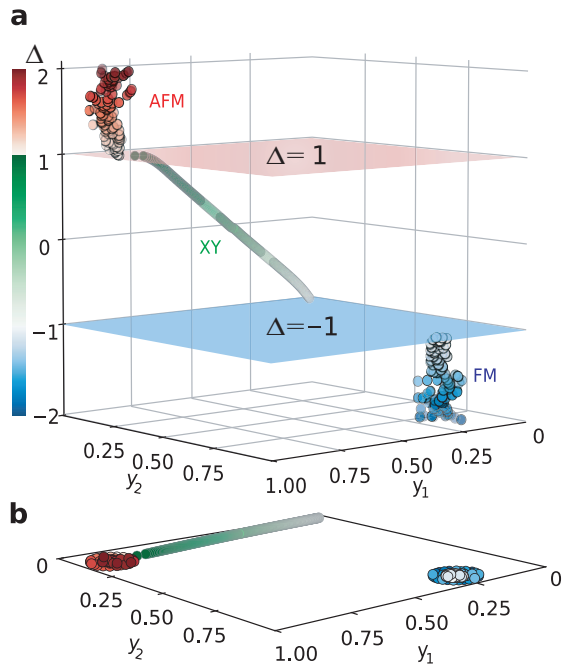}
	\caption{(Color online) {\bf{a}} The three-dimensional (3D) visualization of distinct phases in the 1D anisotropic XXZ antiferromagnetic (AFM) Heisenberg model with anisotropy $\Delta$. In addition to the two dimensions of features $y_1$ and $y_2$ of $\mathcal{R}^2$, $\Delta$ is also plotted as the third dimension for a better visualization. Two expected transition points at $\Delta = -1$ and $1$ are indicated by two semitransparent planes to assist visualization.
{\bf{b}} The visualization of {\bf{a}} in feature space $\mathcal{R}^2$. Three phases (AFM, XY and FM) are clearly visualized with different patterns in the 3D or 2D space. There are $400$ ground states calculated by discretizing $\Delta$ with an interval $0.01$. We take the system size $L=120$, the dimension cut-off of DMRG $\chi=160$, iteration steps in $t$-SNE $n_{iter}=5000$ with the perplexity $\mathcal{P}=24$.}
\label{fig_XXZhalf_3d}
\end{figure}

\subsection{Identifying multiple phases in XXZ spin chain}

Determining the critical points of more than two phases is challenging with the existing machine-learning-based methods such as confusion \cite{van2017learning}. We consider the 1D spin-$\frac{1}{2}$ anisotropic XXZ model $\hat{H}=\sum_{i} (\hat{S}_i^x \hat{S}_{i+1}^x + \hat{S}_i^y S_{i+1}^y) + \Delta \sum_{i} \hat{S}_i^z \hat{S}_{i+1}^z$ \cite{giamarchi2003quantum} with $\Delta$ representing the magnetic anisotropy. This system possesses three phases, say FM ($\Delta < -1$), XY ($-1<\Delta < 1$), and AFM phases ($\Delta> 1$) \cite{giamarchi2003quantum}. Fig.~\ref{fig_XXZhalf_3d} {\bf{a}} shows the visualization of the quantum phases of this model in the space spanned by feature 1, feature 2, and the anisotropy parameter $\Delta$. The expected transition points $\Delta = -1$ and $\Delta = 1$ are indicated by two semitransparent planes. While the states in the FM or AFM phase cluster within the two oval regions of $\mathcal{R}^2$ [see Fig. \ref{fig_XXZhalf_3d} {\bf{b}}], the states in the XY phase form a 1D stream. The phase transition points can be accurately identified as the end points of this stream, which touch on the $\Delta=1$ and $-1$ planes, respectively.
\\
\section{Automatic Identification of Quantum Phases by \emph{k}-means}

\begin{figure*}[htb]
	\includegraphics[width=0.9\linewidth]{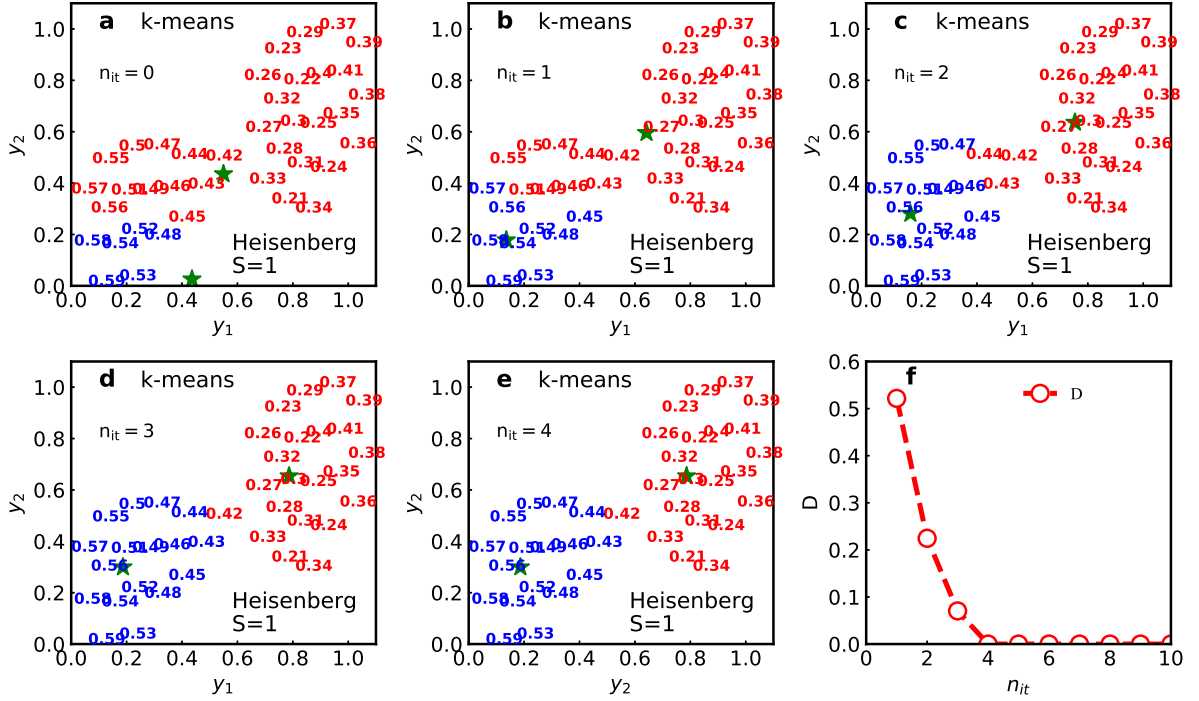}
	\caption{(Color online) {\bf{a}}-{\bf{e}} demonstrate the evolution of samples classified into different clusters (marked by red or blue) and the centers of clusters (indicated by the two stars) after different iteration time in $k$-means. The numbers in {\bf{a}}-{\bf{e}} denote the magnitudes of the applied magnetic fields. {\bf{f}} shows the difference of the centers before and after the $n_{it}$-th iteration [see Eq. (\ref{eq-kD})], showing that the iteration converges only after $n_{it}=4$ steps.}
	\label{fig_Haldane_km}
\end{figure*}

\begin{figure}[htb]
	\includegraphics[width=1\linewidth]{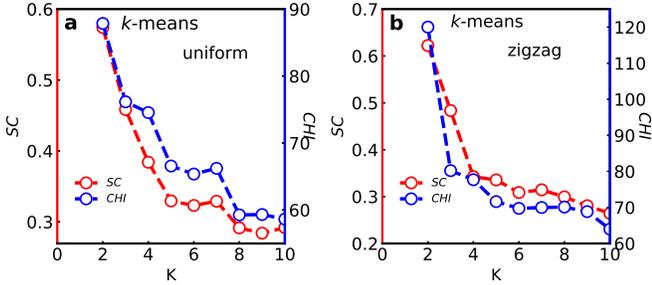}
	\caption{(Color online) The $SC$ [Eq.~(\ref{eq:SC})] and $CHI$ [Eq.~(\ref{eq:CHI})] versus $K$ for the spin $S=1$ antiferromagnetic Heisenberg uniform chain {\bf{a}} and zigzag chain {\bf{b}}.}
	\label{fig_HaZz_km}
\end{figure}

\begin{figure*}[htb]
	\includegraphics[width=0.9\linewidth]{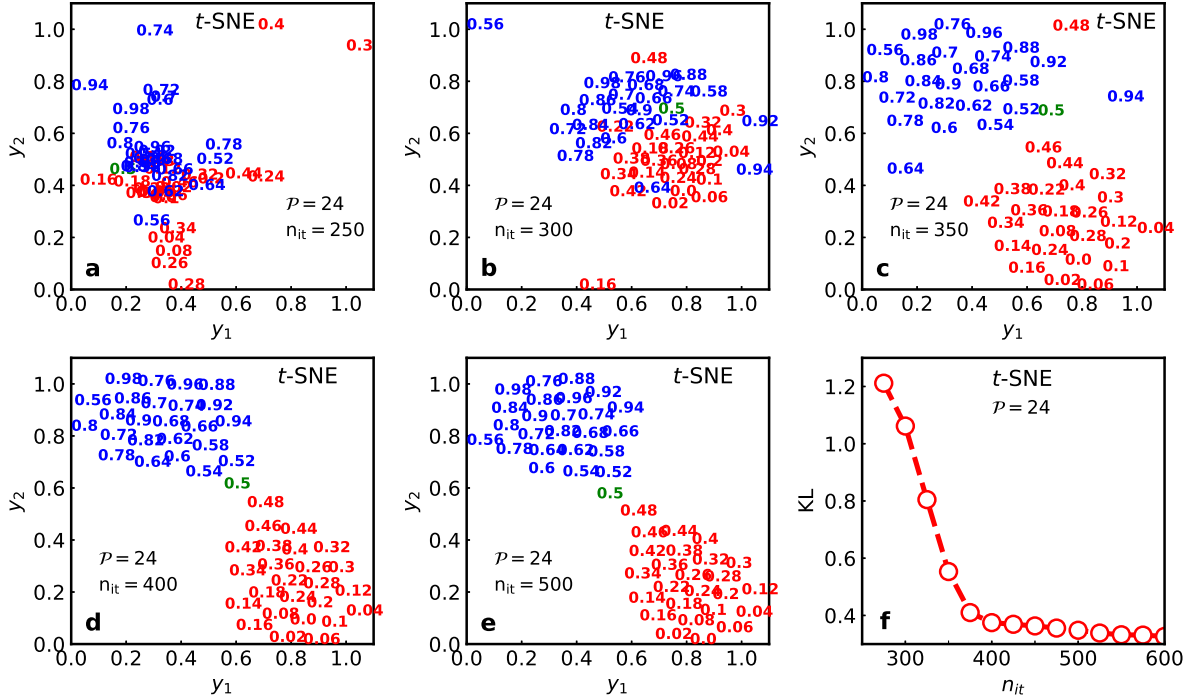}
	\caption{(Color online) {Distribution of quantum states in 2D feature space varies with iteration time of $t$-SNE.} The distribution in $\mathcal{R}^2$ of the grounds states of TFIM by $t$-SNE with the iteration time ${n_{it}}=$250, $300$, $350$ $400$, $500$ in {\bf{a}}-{\bf{e}}, respectively. {\bf{f}} shows the KL divergence versus ${{n_{it}}}$. Here we take the perplexity $\mathcal{P}=24$, the system size $L=80$, and dimension cut-off in DMRG $\chi=30$. The red and blue numbers represent the applied magnetic fields, and the green number $0.5$ denotes the critical field.}
	\label{fig_Ising_niter}
\end{figure*}

\begin{figure*}[tbp]
	\includegraphics[width=0.9\linewidth]{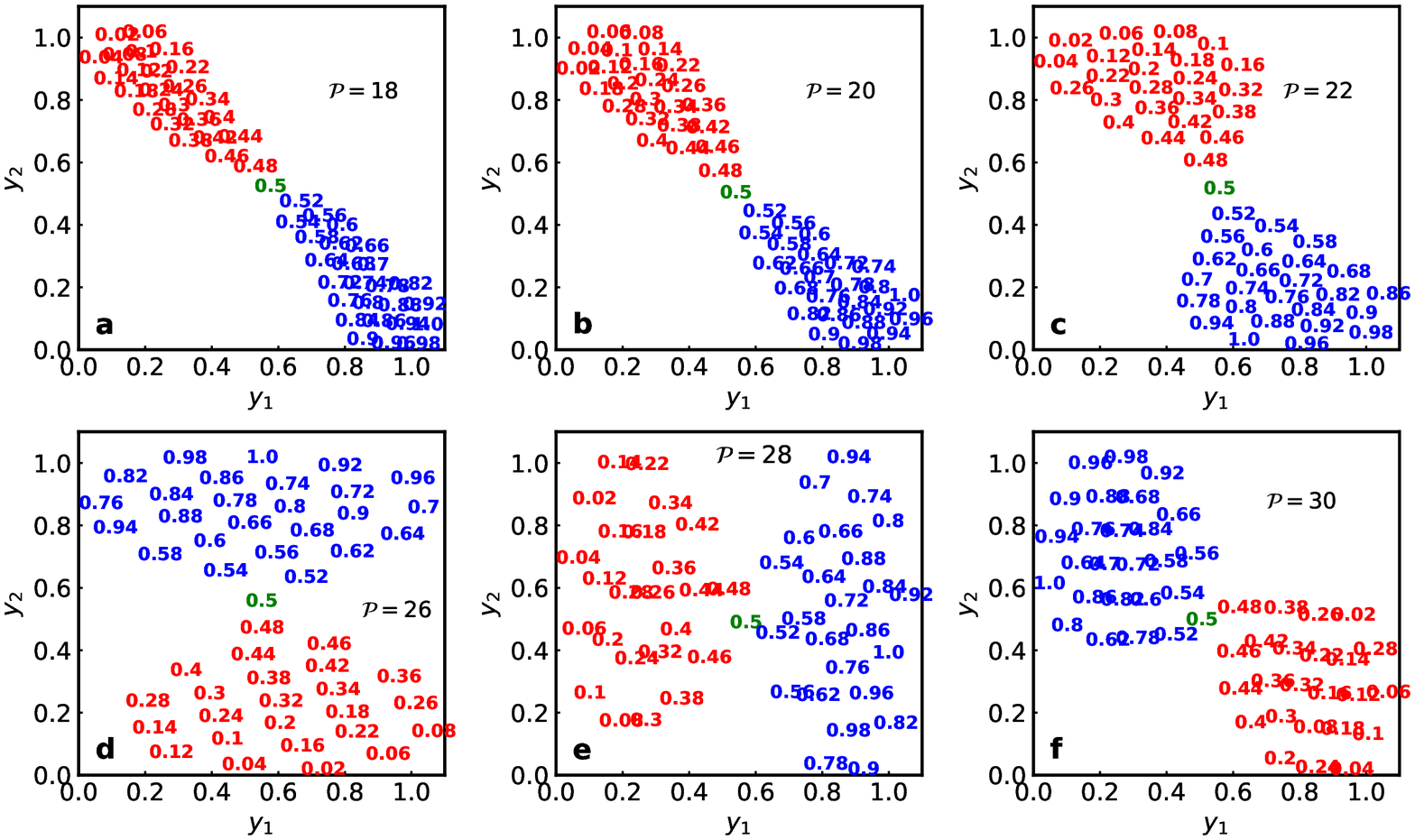}
	\caption{(Color online) {Distribution of quantum states in 2D feature space varies with perplexities of $t$-SNE.} Visualization of the grounds states of TFIM  by $t$-SNE with different perplexities $\mathcal{P}=18$, $20$, $22$, $26$, $28$, and $30$. We take the total iteration times in $t$-SNE ${n_{it}}=5000$, the system size $L=80$, and dimension cut-off in DMRG $\chi=30$.}
	\label{fig_Ising_per}
\end{figure*}

After mapping the ground states to the two-dimensional feature space $\mathcal{R}^2$, we show that different quantum phases can be distinguished simply by naked eyes from how the quantum states are distributed in $\mathcal{R}^2$. Below, we show that one may use $k$-means algorithm \cite{kriegel2011density} to classify the states based on the distributions in $\mathcal{R}^2$.

$K$-means method is an unsupervised learning algorithm and can be used to implement classification tasks. For a set of samples $\left\{ {\mathbf{y}^{\alpha}} \right\}$, $k$-means partitions them into $K$ clusters $\left\{ {{\mathbb{S}^k}} \right\}$ with $k=1, \ldots, K$. The center of each cluster (denoted as $\left\{ {{{\bf{m}}^k}} \right\}$; also called the centroids) can be defined by the samples therein as
\begin{equation}
{\mathbf{m}}^k = \frac{1}{{{N^k}}}\sum\limits_{{{\mathbf{y}}^\alpha } \in {\mathbb{S}^k}} {{{\mathbf{y}}^\alpha }},
\label{eq-k-means2}
\end{equation}
where ${{N^k}}$ is the number of samples in ${\mathbb{S}^k}$.

To classify $\left\{ {\mathbf{y}^{\alpha}} \right\}$, one performs the following two steps iteratively. The first step is to assign the samples to the $K$ clusters according to the given $\left\{ {{{\bf{m}}^k}} \right\}$, where any sample in a given cluster should possess the smallest Euclidean distance to the center of this cluster than to other centers. It means that the samples are divided into $K$ sets $\left\{ {{\mathbb{S}^k}} \right\}$ by satisfying
\begin{equation}
{\mathbb{S}^k} = \left\{ {{{\mathbf{y}}^\alpha }:\left\| {{{\mathbf{y}}^\alpha } - {{\mathbf{m}}^k}} \right\| \leqslant \left\| {{{\mathbf{y}}^\alpha } - {{\mathbf{m}}^j}} \right\| \ \forall j} \right\},
\label{eq-k-means1}
\end{equation}
where ${\left\| {{{\mathbf{y}}^\alpha } - {{\mathbf{m}}^k}} \right\|}$ represents the Euclid distance between the sample ${{{\mathbf{y}}^\alpha }}$ and the center ${{{\mathbf{m}}^k}}$, and $j$ goes over all centers. The second step is to update $\left\{ {{\mathbf{m}}^k} \right\}$ based on the present $\left\{ {{\mathbb{S}^k}} \right\}$ according to Eq. (\ref{eq-k-means2}). These two steps are executed iteratively until $\left\{ {{\mathbf{m}}^k} \right\}$ converges.

We apply $k$-means to categorize the ground states of spin-1 antiferromagnetic Heisenberg chain into two phases ($K=2$) after mapping those states onto $\mathcal{R}^2$ space by $t$-SNE. In Fig. \ref{fig_Haldane_km}, the numbers represent the ground states with reduced dimensionality $\{{{\mathbf{y}}^\alpha }\}$ in different magnetic fields $\alpha$, and the two stars represent the centers $\left\{ {{\mathbf{m}}^k} \right\}$. The states divided into two clusters are marked by different colors. To begin with, one first randomly initializes the positions of the centers, with which the states are divided into two clusters according to Eq. (\ref{eq-k-means1}). After four steps of iterations, $\left\{ {{\mathbf{m}}^k} \right\}$ converges, and the states in different phases are successfully divided to the two clusters. Fig. \ref{fig_Haldane_km} {\bf{f}} shows how the centers converge by
making use of
\begin{equation}
D(n_{it}) = \sum\limits_k^K {\left\| {{\mathbf{m}}^k(n_{it}) - {{\mathbf{m}}^k(n_{it}-1)}} \right\|},
\label{eq-kD}
\end{equation}
with $\left\{ {{\mathbf{m}}^k(t)} \right\}$ the centers after $t$ iterations. We find that $D(t)$ decreases almost to $0$ for $n_{it}=4$.

Though $K$ is previously known in the above example, it can also be determined automatically when one does not know how many clusters that the samples should be divided into. We refer to the silhouette coefficient ($SC$) \cite{rousseeuw1987silhouettes} and Calinski-Harabasz index ($CHI$) \cite{maulik2002performance} for this purpose. The $SC$ is defined as
\begin{subequations}
\begin{align}
& SC = \frac{1}{J}\sum\limits_{\alpha  = 1}^J {\frac{{\overline {D_{out}^\alpha }  - \overline {D_{in}^\alpha } }}{{max(\overline {D_{out}^\alpha } ,\overline {D_{in}^\alpha } )}}}, \label{eq:SC} \\
& \overline {D_{in}^\alpha }  = \frac{1}{{{N^{_{k\left( \alpha  \right)}}} - 1}}\sum\limits_{{{\mathbf{y}}^{\alpha '}} \in {\mathbb{S}^{k\left( \alpha  \right)}}} {\left\| {{{\mathbf{y}}^{\alpha '}} - {{\mathbf{y}}^\alpha }} \right\|}, \\
& \overline {D_{out}^\alpha }   = \frac{1}{{N - {N^{_{k\left( \alpha  \right)}}}}}\sum\limits_{{{\mathbf{y}}^{\alpha '}} \notin {\mathbb{S}^{k\left( \alpha  \right)}}} {\left\| {{{\mathbf{y}}^{\alpha '}} - {{\mathbf{y}}^\alpha }} \right\|} ,
\end{align}
\end{subequations}
where ${k\left( \alpha \right)}$ represents the cluster that the $\alpha$-th data point belongs to, $\overline{D_{in}^j}$ ($\overline{D_{out}^j}$) is the average distance of sample $j$ to others in (not in) the same cluster. The value of $SC$ ranges from -1 to 1. The optimal $K$ is chosen so that $SC \to 1$ \cite{rousseeuw1987silhouettes}.

The $CHI$ is defined as \cite{maulik2002performance}:
\begin{subequations}
\begin{align}
& CHI = \frac{Trace(B)/(K-1)}{Trace(W)/(J-K)}, \label{eq:CHI} \\
& Trace(B) = \sum\limits_{k = 1}^K {{N^k}} \parallel {{\mathbf{m}}^k} - {{\mathbf{m}}^0}{\parallel ^2}, \\
& Trace(W) = \sum\limits_{k = 1}^K {\sum\limits_{{{\mathbf{y}}^\alpha } \in {\mathbb{S}^{k\left( \alpha  \right)}}} {{{\left\| {{{\mathbf{y}}^\alpha } - {{\mathbf{m}}^k}} \right\|}^2}} }.
\end{align}
\end{subequations}
$B$ is the between-cluster scatter matrix and W is the within-cluster scatter matrix; ${{\mathbf{m}}^0}$ is the centroid of the whole dataset. The optimal $K$ is chosen so that $CHI$ reaches its maximum. \cite{maulik2002performance}.

For the grounds states of the spin $S=1$ antiferromagnetic Heisenberg uniform chain and zigzag chain, Fig. \ref{fig_HaZz_km} shows the $SC$ and $CHI$ calculated from the distribution of the ground states in $\mathcal{R}^2$. One can see that the optimal number of clusters should be $K=2$, consistent with the fact that there are two phases for each system. In this way, one does not need priori knowledge about either the properties of the original states or the number of phases.



\begin{figure*}[htp]
	\includegraphics[width=0.9\linewidth]{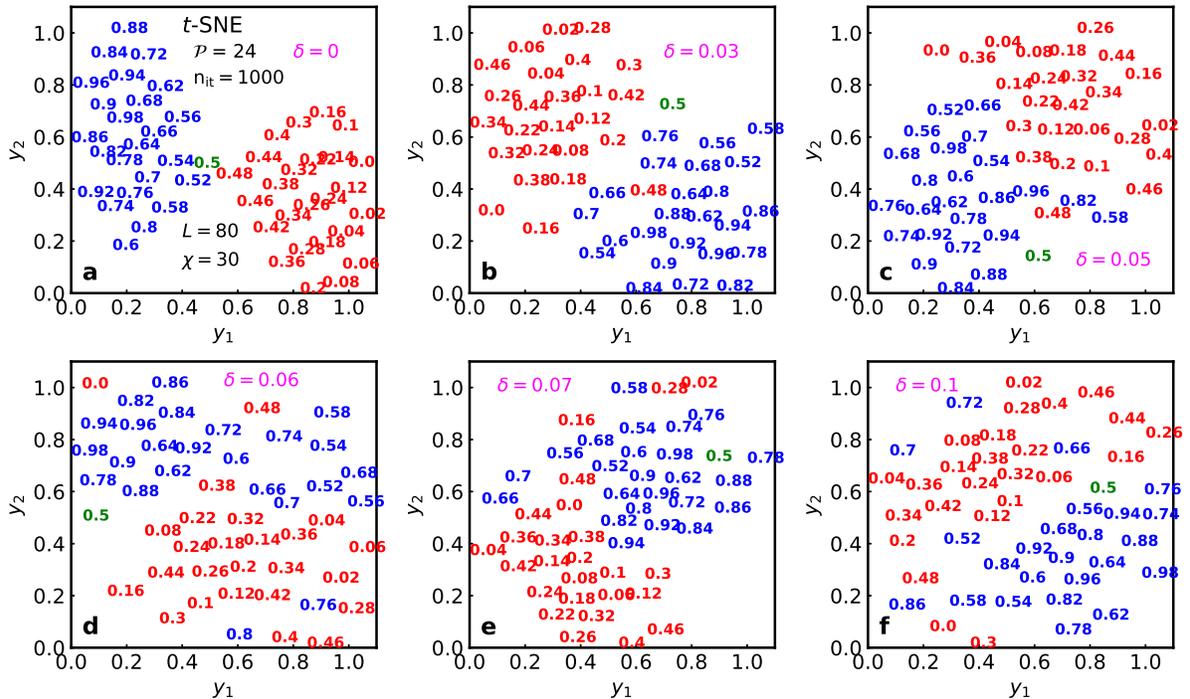}
	\caption{(Color online) {Distribution of quantum states in 2D feature space varies with noises.} The visualization of the ground states of TFIM under small noise by $t$-SNE with different strengths of noise $\delta=$0, 0.03, 0.05, 0.06, 0.07, and 0.10 in {\bf{a}}-{\bf{f}}, respectively. We take the perplexity $\mathcal{P}=24$, the total iteration time ${n_{it}}=1000$, the system size $L=80$, and dimension cut-off in DMRG $\chi=30$. }
	\label{fig_Ising_delta}
\end{figure*}

\begin{figure}[htb]
	\includegraphics[width=0.9\linewidth]{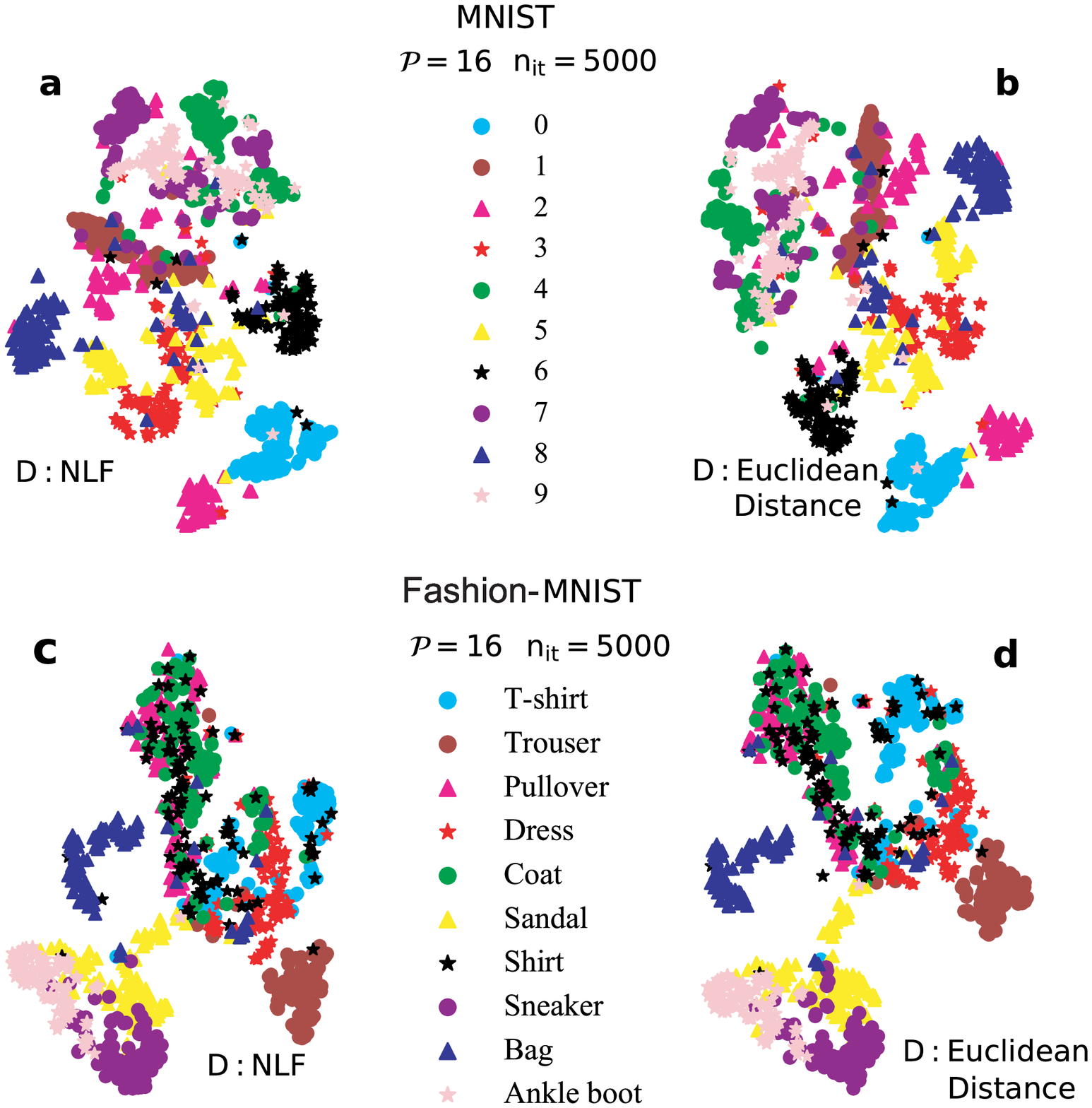}
	\caption{(Color online) {Visualization of classical datasets by $t$-SNE with NLF and Euclidean distance.} The visualization of the MNIST dataset and the fashion-MNIST dataset by $t$-SNE with NLF or the Euclidean distance. In {\bf{a}} and {\bf{c}}, the images are mapped onto the Hilbert space, where the distance of two different images are measured by NLF.  In {\bf{b}} and {\bf{d}}, the distance of two different images are measured by the Euclidean distance in the original feature space. For both MNIST and Fashion-MNIST datasets, we take 1000 images (100 images from each class) as the input of $t$-SNE. We take $\mathcal{P}=16$ and total iteration times ${n_{it}}=5000$.}
	\label{fig_mnist_fashion}
\end{figure}

\section{Robustness of $t$-SNE for quantum phase identification}
\label{sec::robust}
Fig.~\ref{fig_Ising_niter} show in $t$-SNE, how the distribution in $\mathcal{R}^2$ of the ground states of the transverse field Ising model converges. Remind that for $N$ given states $|\psi^{\alpha}\rangle$ in the Hilbert space $\mathcal{H}$, the $t$-SNE directly optimizes $\mathbf{y}^{\alpha}$ in $\mathcal{R}^2$ that are the $N$ corresponding low-dimensional vectors after reducing the dimensionality. Initially, $\mathbf{y}^{\alpha}$ are randomly determined. In Fig.~\ref{fig_Ising_niter} {\bf{a}}-{\bf{e}} show the $\mathbf{y}^{\alpha}$ after ${n_{it}}$ iterations with ${n_{it}}=250, 300, 350, 400, 500$, respectively. One can see that after  ${n_{it}} \simeq 400$ iterations, the distribution converges, where the two quantum phases are clearly visualized. The KL divergence, which indicates the difference between the distributions of the samples in $\mathcal{H}$ and $\mathcal{R}^2$, decays with ${n_{it}}$ as shown in Fig.~\ref{fig_Ising_niter}~{\bf{f}}.
A GIF was provided \cite{gif} to animatedly show how the states cluster in $\mathcal{R}^2$ as the iteration time increases.

From the previous works in machine learning, it is known that the visualization by $t$-SNE is robust to the perplexity $\mathcal{P}$. Fig.~\ref{fig_Ising_per} shows that the ground states of the transverse field Ising model (TFIM) by $t$-SNE with different perplexities form similar hourglass-like patterns. The difference of these patterns is to what extent the hourglass extends in the two-dimensional plane. This is consistent with the fact that the perplexity controls how nonlocally one state is correlated to others from the joint probability distribution. More specifically, as the dimensionality is reduced, the distribution in $\mathcal{R}^2$ may not respect the mutual relations among the states in $\mathcal{H}$. For instance, it is possible that one has $||\mathbf{y}^1-\mathbf{y}^2|| < ||\mathbf{y}^1-\mathbf{y}^3||$ in $\mathcal{H}$ but $||\mathbf{y}^1-\mathbf{y}^2|| > ||\mathbf{y}^1-\mathbf{y}^3||$ in $\mathcal{R}^2$. A small perplexity means that the distribution in $\mathcal{R}^2$ should in prior satisfies the mutual relations of distances for those with small distances. Consequently in the visualization with small $\mathcal{P}$, different clusters tend to separate apart mutually. This leads to a ``thinner'' hourglass than those with larger perplexities. Note that in practice, the perplexity is usually smaller than the number of samples \cite{maaten2008visualizing}.

We also investigate the visualization of quantum phases under noises. The noisy quantum states are defined by
\begin{equation}
|\Psi_{\delta}^{\alpha}\rangle = \sqrt{1-\delta}|\Psi^{\alpha}\rangle + \sqrt{\delta}|\Psi_{random}\rangle.
\label{eq::noise_state}
\end{equation}
where $|\Psi^{\alpha}\rangle$ is the ground state in the transverse magnetic field $\alpha$, and $\delta$ is a small constant that controls the strength of the noise. $|\Psi_{random}\rangle$ is a random matrix product state (MPS) whose bond dimensions are identical to those of the ground states. All elements of the tensors in $|\Psi_{random}\rangle$ are generated randomly by the Gaussian distribution $N(0, 1)$.

In Fig.~\ref{fig_Ising_delta}, we show the visualizations of $\{|\Psi^{h_x}_{\delta} \rangle \}$ with $\delta$= 0, 0.03, 0.05, 0.06, 0.07 and 0.10. For $\delta\leq0.03$, the clusters of the AFM and FM phases are clearly separated, and the boundary of the clusters successfully gives the critical point. By increasing $\delta$ to $\delta>0.03$, two clusters gradually merge into each other, and it becomes more and more difficult to identify the critical point. These results suggest that our quantum phase visualization scheme is robust against small random noises.

\section{Visualization of classical data with negative logarithmic fidelity}

Below, we show that our scheme can also be applied to visualize classical data, such as the images in the MNIST \cite{6296535} and fashion-MNIST \cite{xiao2017fashion} datasets. To calculate the negative logarithmic fidelity (NLF) of the classical samples, we firstly map each pixel $x^\alpha_n$ to the Hilbert space \cite{stoudenmire2016supervised} by
\begin{equation}
|\phi ({x^{\alpha}_n})\rangle = \cos\frac{\pi {x}^{\alpha}_n}{4} |0\rangle + \sin\frac{\pi {x}^{\alpha}_n}{4} |1\rangle,
\label{eq:feature_map}
\end{equation}
where ${{\mathbf{x}}^{\alpha}_n}$ with $0 \leq \mathbf{x}_n \leq 1$ denotes the value of the $n$-th pixel in the $\alpha$-th image, and $\{|i\rangle\}$ ($i=0, 1$) denote the orthonormal basis in the two-dimensional Hilbert space. Then an image can be mapped to a product state as
\begin{equation}
|\psi^{\alpha} \rangle = |\phi ({{x}_1^{\alpha}}) \rangle \otimes |\phi ({{x}_2^{\alpha}}) \rangle \otimes  \cdots |\phi ({{x}_n^{\alpha}}) \rangle,
\label{eq:product_state}
\end{equation}
Obviously, $|\psi^{\alpha} \rangle$ is a state defined in the $2^L$-dimensional Hilbert space with $L$ the total number of pixels in one image. The NLF between two images is defined as
\begin{equation}
D_{\alpha, \alpha '}^{\cal H} =  - \log \left( {\left\langle {{\psi^\alpha }} \right|\left. {{\psi^{\alpha '}}} \right\rangle } \right).
\label{eq:product_state}
\end{equation}
With $D_{\alpha, \alpha '}^{\cal H}$, the images can be visualized by $t$-SNE by following the same the steps for visualizing the ground states.

The visualizations of the images in MNIST and fashion-MNIST based on NLF are shown in Fig. \ref{fig_mnist_fashion} {\bf{a}} and {\bf{c}},  respectively. As a comparison, the visualizations using the Euclidean distance $D_{\alpha, \alpha '}^{\cal E} = \| \mathbf{x}^{\alpha} - \mathbf{x}^{\alpha'} \|$ in the $t$-SNE are shown in Fig. \ref{fig_mnist_fashion} {\bf{b}} and {\bf{d}} for MNIST and fashion-MNIST, respectively. Both schemes show similar visualization results, which indicates that our scheme also works well for visualizing classical data.

\section{Discussion And Conclusion}
Our results show that the states in the XY phase, which are critical and can be described by the conformal field theory with central charge $c=1$, forming a 1D stream in $\mathcal{R}^2$. In contrast, the non-critical phases (the FM/AFM and gapped topological phases) exhibit oval clusters. To explain the cause of different patterns of the distributions given by the critical and non-critical phases, we propose the following intuitive arguments. As the distance of two states (in both $\mathcal{H}$ and $\mathcal{R}^2$) is positively associated with the difference of their physical quantities (e.g., magnetizations, correlations, entanglement spectrum, etc.), the states within each phase should cluster because they share similar physics and thus should have small distances between each other.

The physics of the states within the gapped phase are almost identical. The distances among the states within each phase are insignificant. Even the energy levels do cross due to finite-size effects or numerical errors, the differences of the physics for these states should be minor. Therefore, the quantum states are expected to cluster in a small region in $\mathcal{R}^2$. The situations for the gapless but non-critical phases are similar. Take the polarized phase in TFIM as an example. For different $h_x$'s with $h_x>0.5$, the distances among the states are more minor than the distances between the states in different phases. In the vicinity of the critical point, the gap gradually closes, and the energy levels become dense, implying that the physical properties change more drastically as the physical parameter alters. It turns out that the distances between the quantum states with different parameters become more significant in this region than those within the non-critical phases.

When two non-critical phases are separated by a critical phase instead of a critical point, the distances of the states within the critical phase should be more significant than those within the non-critical phases. This leads to the distribution of the GS's of XXZ model (Fig.~\ref{fig_XXZhalf_3d}) in the critical phase forms a 1D stream in $\mathcal{R}^2$.

In conclusion, we propose a scheme to visualize quantum phases and to identify phase transition points via machine learning. The key idea is to map the quantum states in Hilbert space $\mathcal{H}$ where the distribution of ground states is difficult to access onto the 2D feature space $\mathcal{R}^2$ by the nonlinear DR method $t$-SNE, where the negative logarithmic fidelity is adopted to measure the distances between different quantum states. It is found that the distribution in $\mathcal{R}^2$ exhibits different patterns for distinct phases, from which the phase transition points can be readily identified. The success of this proposal is demonstrated on a few of 1D quantum many-body models, including those with conventional phases within Landau paradigm, the topological phases with nonlocal orders, and the critical phase described by CFT.
This present strategy for visualization through learning works well not only for quantum data but also for classical data.

While our scheme of visualizing quantum phases via learning are flexible and general, more rigorous and robust relations between the distributions in $\mathcal{R}^2$ and the physical properties of the quantum phases (e.g., criticality and topology) space are to be established. As a non-linear DR method, the $t$-SNE works as a ``black box'' which  guarantees the minimization of the KL-divergence in a variational sense, and it is unknown how to interpret it, for instance, what the two features ($y_1$ and $y_2$ in $\mathcal{R}^2$) stand for. It is really interesting to seek for the DR methods with higher interpretability, which would assist us to unveil more novel properties of quantum many-body systems by this visualization scheme.
\\

\begin{acknowledgments}
This work is supported in part by the NSFC (Grant No. 11834014), the National Key R\&D Program of China (Grant No. 2018FYA0305804),  the Strategetic Priority Research Program of the Chinese Academy of Sciences (Grant No. XDB28000000), and Beijing Municipal Science and Technology Commission (Grant No. Z190011).
SJR is supported by Beijing Natural Science Foundation (No. 1192005 and No. Z180013), Foundation of Beijing Education Committees (No. KM202010028013), and the Academy for Multidisciplinary Studies, Capital Normal University.
\end{acknowledgments}	

%

\end{document}